\documentclass[twocolumn,prl,amsmath,amssymb,aps,floatfix,longbibliography]{revtex4-2}
\usepackage{xcolor}

\usepackage{graphicx}
\usepackage{dcolumn}
\usepackage{bm}

\begin{document}

\preprint{APS/123-QED}

\title{Robust Edge States from Band Topology in a Damped One-dimensional Magnonic Crystal}

\author{Kwangyul Hu}
 \email{kwangyul-hu@uiowa.edu}
 \affiliation{Department of Physics and Astronomy, University of Iowa, Iowa City, Iowa 52242,
USA}
\author{Denis R. Candido}
 \email{denis-candido@uiowa.edu}
\affiliation{Department of Physics and Astronomy, University of Iowa, Iowa City, Iowa 52242,
USA}
\author{Michael E. Flatt\'e}%
 \email{michaelflatte@quantumsci.net}
\affiliation{Department of Physics and Astronomy, University of Iowa, Iowa City, Iowa 52242,
USA}
\affiliation{Department of Applied Physics, Eindhoven University of Technology, P.O. Box 513, 5600 MB, Eindhoven, The Netherlands}

\date{\today}

\begin{abstract}

The presence or absence of topologically-produced edge states of a crystal are robust to disorder; their stability in the presence of decay is less clear. For topologically nontrivial bosonic systems with finite particle lifetimes, such as photonic, phononic, or magnonic structures, a natural hypothesis suggests that if the linewidth from particle decay exceeds the gap between neighboring bands, then topological features such as Berry phases or edge states will lose their protection. Here we show that topological properties are significantly more robust than this, by assessing the properties of a one-dimensional magnonic crystal as the damping is increased. Even when the damping greatly exceeds the gap between neighboring bands the Zak phase of those bands is nearly unchanged, and the edge states remain clearly visible in micromagnetic simulations of microwave transmission.  These results clarify the understanding of robust topological properties and  bulk-boundary correspondence.

\end{abstract}

\maketitle

Studies of band topology in one-dimensional (1D) solid state systems provide rich topological features within a simpler setting than  higher dimensional topological insulators. A central topological invariant in 1D ideal crystals is a Berry phase also known as the Zak phase\cite{ZakP_1}. For a system with inversion symmetry, the Zak phase is quantized: either $0$ or $\pi$. Topologically protected interface modes at a boundary of two 1D crystals with different Zak phases have been identified in  phononic\cite{Top1D_1,Top1D_2,Top1D_3}, photonic\cite{Top1D_4,Top1D_5}, and magnonic crystals\cite{TopMag_9} and also metamaterials\cite{Top1D_6,Top1D_7}. Nevertheless, there is only limited discussion about the topological characteristics of an individual magnonic system similar to the Su-Schrieffer-Heeger(SSH) model\cite{SSH_1,SSH_2,SSH_3} in the presence of damping. Magnonic systems constructed from some materials such as yttrium iron garnet (YIG) 
possess  very low losses ({\it e.g.} quality factors greater than 10$^4$) at microwave frequencies. Interface modes robust to damping in magnonic structures may, {\it e.g.}, permit high-bandwidth lossless long-range information transmission on chip-relevant length scales ($\sim 100$'s~$\mu$m), and also enable fundamental studies of switchable topological properties through variation of a global external field. 

Here we characterize the band topology of an isolated 1D magnonic crystal (MC) polarized with an external magnetic field (bias field) normal to the thin-film surface, and identify a surprising robustness of the topological invariant even when damping closes the gap. We identify the topological features of inversion-symmetric MCs composed of alternating layers, along a spatial direction $y$,  of different magnetic material using magnonic band structures calculated from the Landau-Lifshitz-Gilbert(LLG) equation\cite{MagC_1,MagC_2,MagC_3,MagC_4,MagC_5,MagC_6,MagC_7,MagC_8,MagC_9,TopMag_2}. 
In 1D electronic systems, the electron wave functions are used to analyze the edge states\cite{ZakP_3}. Such an analysis based on magnon wave functions is impossible for a 1D MC due to the absence of magnetization outside of the MC. We address this fundamental issue using a dynamical property present both inside and outside the MC: the dynamic (dipolar) magnetic field. The different Zak phases of bands coincide with a localization of this  magnetic field at the edges, which also reflects confinement of the dynamic magnetization within the MC near the interface and the existence of topologically protected edge states. In this system the band edges occur either at crystal momenta $k=0$ or $k={\pi}/{a}$. {In the absence of damping we} connect the spatial parity of the band edge magnetic field $\beta_{\ell,k}(y)$
with the Zak phase $\phi_\ell$
of  band $\ell$, and to the sign of the field's {logarithmic derivative} $\rho(k,y) = d \ln \beta_k(y) /dy$ within the nearest band gap. For a negative logarithmic derivative, the resulting magnetic field and wave function have a complex wave number $k$ and describe an edge state in the band gap. To compare with these analytic results the topologically protected edge states are directly identified by driving bulk and finite MCs with a microwave magnetic field in micromagnetic LLG simulations that include damping. Edge state excitations in asymmetric MCs show surprising results; 
an edge state in the first gap separates into two edge states that are localized at each end of the MC, however the edge state of the second gap becomes locked to one termination of the MC with a shifted frequency. Finally, we calculate the Zak phases and dispersion relations for larger Gilbert damping constants and show that the topological properties of the MC are robust even for damping large enough to close the gaps of the MC.

\begin{figure}
    \centering
    \includegraphics{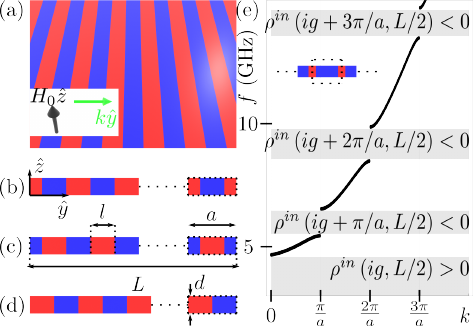}
    \caption{Schematic MC with lattice constant $a$ and thickness $d$; $l$ is the width of the red slab and ${l}/{a}$  the ``filling fraction''. (a) 1D 
    infinite  MC with bias field (black arrow) and spin wave propagation direction (green arrow). (b-d) Side view of  finite structures: (b) red slab termination, (c) blue slab termination, (d) asymmetric  termination. (e) Band structure of an infinite MC  [(a)], 
    showing the sign of $\rho^{in}$ 
    within different band gaps. 
    }
    \label{fig1}
\end{figure}

A schematic 1D MC is illustrated in Fig~\ref{fig1}, consisting of periodically arranged blue and red slabs representing two different magnetic materials, where the blue slab has a lower saturation magnetization, $M_{s,b}$,  and exchange stiffness constant, $A_b$, than the red ($M_{s,r}$, $A_r$). Our specific example here explores alternating (red) yttrium iron garnet ($\text{YIG}$) and (blue) yttrium aluminum iron garnet ($\text{Y}_{3}\text{Al}_{x}\text{Fe}_{5-x}\text{O}_{12}$, $\text{YAIG}$). 
Substitution of nonmagnetic  $\text{Al}$ for $\text{Fe}$ reduces the magnetism  of the material\cite{YAIG_1,YAIG_2,YAIG_3}, and for the calculations here the $\text{YAIG}$ has $x\approx0.38$, corresponding to an $M_s$ and $A$ that are $60\%$ of YIG's $M_s=1.47\times 10^{5}$ $\text{A}/\text{m}$ and $A=4.09\times10^{-12}$.
Figure~\ref{fig1}(b-d) shows three different MC terminations for the finite structure analysis; (b) and (c) have inversion symmetry but (d) does not. As the Zak phase formalism requires inversion symmetry a quantization of the Zak phase is expected for {Figs.~\ref{fig1}(b) and (c).}


Fig.~\ref{fig1}(e) shows the  magnonic dispersion and $\rho$, simulated as described below, for an MC lattice constant of $a=128$ $\text{nm}$ and $l/a = 0.25$ along  $\hat y$.  
The thickness of the slab ($\hat z$) is $d=16$~$\text{nm}$, with vacuum at the top and bottom. To focus on the one dimensionality of the structure periodic boundary conditions are used in the $\hat x$ direction and only the lowest width mode quantum number is considered.
The external static field  is $H_{0}\hat z$ with $H_{0}=0.3\text{T}$. 
The magnonic dynamics is simulated with a  linearized LLG equation\cite{MagC_1,MagC_2,MagC_3,MagC_4,MagC_5,MagC_6,MagC_7,MagC_8,MagC_9,TopMag_2},
\begin{eqnarray}
\frac{d}{dt}\textbf{M}=-\gamma\mu_{0}\textbf{M}\times\textbf{H}_{eff}+\frac{\alpha}{M_s}\textbf{M}\times\frac{\partial}{\partial t}\textbf{M},
\end{eqnarray}
where $\textbf{M}=\textbf{m}_{x,y}+M_s\hat{z}$ is the magnetization, $\textbf{H}_{eff}$ the effective magnetic field, $\gamma$ the gyromagnetic ratio, $\mu_0$ the vacuum permeability, and $\alpha$ the Gilbert damping parameter. 
The magnetization dynamics is linear in $m_{x,y}$ when $m_{x,y}\ll M_s$. The effective magnetic field $\textbf{H}_{eff}$ is
\begin{eqnarray}
\textbf{H}_{eff}=H_{0}\hat{z}+H_{d}\hat{z}+\textbf{h}+\textbf{H}_{ex},
\end{eqnarray}
where 
$H_{d}\hat{z}$ is the static dipolar field, $\textbf{h}$ is the dynamic dipolar field and $\textbf{H}_{ex}$ is the exchange field. $H_{d}\hat{z}$ and $\textbf{h}$ are associated with the magnetization $M_s\hat{z}$ and $\textbf{m}_{x,y}$ through the magnetostatic Maxwell's equations $\nabla\times\textbf{H}=0$ and $\nabla\cdot\left(\textbf{H}+\textbf{M}\right)=0$ with $\textbf{H}=H_d\hat{z}+\textbf{h}$. The exchange field\cite{Exch} $\textbf{H}_{ex}=\left(\nabla\cdot\lambda_{ex}^{2}\left(\textbf{r}\right)\nabla\right)\textbf{M}$, where $\lambda_{ex}^{2}={2A}/{\mu_{0}M_{s}^{2}}$.

The  LLG equation then has the form
\begin{eqnarray}
i\Omega m_{x}+i\Omega \alpha m_{y}=&&m_{y}H_{eff,z}-M_sh_y,\nonumber\\
i\Omega m_{y}-i\Omega \alpha m_{x}=&&-m_{y}H_{eff,z}-M_sh_x,
\end{eqnarray}
where $\Omega={\omega}/{\gamma\mu_0}$. $m_{x}$ and $m_{y}$, as well as the magnetic parameters $M_s$, $\lambda_{ex}$ and $\alpha$, are expressed as truncated Fourier series of $N$ terms. 
For the unit cell in Fig.~\ref{fig1}(b), $M_s\left(G_{n}\right)$ is 
\begin{eqnarray}
M_s\left(G_{n}\right)=
\begin{cases}
M_{s,r}+M_{s,b}[1-({l}/{a})] & G_{n}=0,\\
\left(M_{s,r}-M_{s,b}\right)\left[(l/a){{\rm sinc}\left({ln\pi}/{a}\right)}\right] & G_{n}\neq 0,
\end{cases}
\end{eqnarray}
where $G_n={2\pi n}/{a}$ are the reciprocal lattice vectors. 

The resulting\cite{Demag} dipolar fields $H_{d,z}$, and $h_{y}$ are
\begin{eqnarray}
H_{d,z}=\sum_{n=-N}^NM_{s}\left(G_{n}\right)e^{iG_{n}y}\zeta\left(G_{n},d\right),
\end{eqnarray}
\begin{eqnarray}
h_{y}=\sum_{n=-N}^Nm_{y}\left(G_{n}\right)e^{i\left(k+G_{n}\right)y}\left[\zeta\left(k+G_{n},d\right)-1\right],
\end{eqnarray}
where, after averaging over the MC thickness,
\begin{eqnarray}
\zeta\left(G_{n},d\right)=\frac{2\sinh\left(\left|G_{n}\right|d/2\right)}{\left|G_{n}\right|d}e^{-\left|G_{n}\right|d/2},\label{xi}
\end{eqnarray}
and $h_x=0$. The eigenvalues of this finite matrix LLG equation\cite{MagC_1,MagC_2,MagC_3,MagC_4,MagC_5} yield the magnon dispersion relations; the eigenvectors are the corresponding spin wave amplitudes.

The magnetic field $\vec{\beta}=\textbf{h}+\textbf{m}$ 
of the $\ell$-th band,
\begin{eqnarray}
    \left|\beta_{y,\ell,k}\right>=\sum_{n=-N}^{N}m_{y}\left(G_n\right)e^{i\left(k+G_n\right)y}\zeta\left(k+G_{n},d\right),
\end{eqnarray}
where $m_y\left(G_n\right)$ are the $y$-component of eigenvectors of the $\ell$-th band. 
A discrete formulation of the Zak phase\cite{Top1D_2,Top1D_3,Top1D_4,Top1D_5,Top1D_6,Top1D_7,ZakP_2} of the $\ell$-th band is
\begin{eqnarray}
\phi_{\ell}=-\text{Im}\ln\prod_{j=0}^{\mu-1}\left<\beta_{y,\ell,k_j}\right|\left.\beta_{y,\ell,k_{j+1}} \right>,\label{zakphase}
\end{eqnarray}
where $k_j=-({\pi}/{a})+({2j\pi}/{\mu a})$ is the $j$-th component of $\mu$ discretized $k$ points within the first Brillouin zone $\left[-{\pi}/{a},{\pi}/{a}\right)$,  and $\left|\beta_{y,\ell,k_\mu}\right\rangle=e^{-i2\pi y/a}\left|\beta_{y,\ell,k_0}\right\rangle$. Our calculation uses $\mu=30$ and $N=16$.
The Zak phase depends on the parities of the magnetic fields at $k=0$ and $k={\pi}/{a}$ with respect to a symmetry point $y_0$; if they have opposite parity the Zak phase $\phi_\ell = \pi$, if the same parity $\phi_\ell=0$. Note that the Zak phases and parities calculated using either the magnetization and the magnetic field are identical.
%
%
%
%


The presence of localized magnetic field is determined\cite{ZakP_3} by the sign of $\rho\left(k,y=\pm L/2\right)$,
where $k=u+ig$ is complex ($g,u\in\mathbb{R}$). The magnetic field inside the 1D crystal is a Bloch function $\left|\beta^{in}_{k}\right>\propto e^{iky}$ and outside it is a decaying function $\left|\beta^{out}_{k}\right>\propto e^{-\tau \left[y-({L}/{2})\right]}$, so $\rho^{out}\left(k,{L}/{2}\right)<0$. As the logarithmic derivatives must match at the termination point, {\it e.g.}  $y=L/2$, $\rho^{in}\left(k,{L}/{2}\right)<0$ is required for field confinement within a band gap. 
$\left|\beta^{in}_{k}\right>$ decays into the bulk at $L/2$ (is a localized state) only if $g\neq 0$. 
%
The relation between the Zak phase and the sign of $\rho^{in}\left(k,{L}/{2}\right)$ in the MC is summarized\cite{TopMag_9}:
\begin{eqnarray}
\text{sgn}\left[\rho^{in}\left(ig+\frac{\ell\pi}{a},\frac{L}{2}\right)\right]=\prod^{\ell}_{j=0}{\left(e^{i\phi_{j}}\right)},
\label{Zakrho}\end{eqnarray}
where {$\ell$} is the number of bands below the band gap represented by $\rho^{in}\left(ig+{\ell\pi}/{a},{L}/{2}\right)$ {and $\phi_0=0$.} We use Eq.~(\ref{Zakrho}) to determine whether topological spin wave edge states are present.

\begin{figure}
    \centering
    \includegraphics{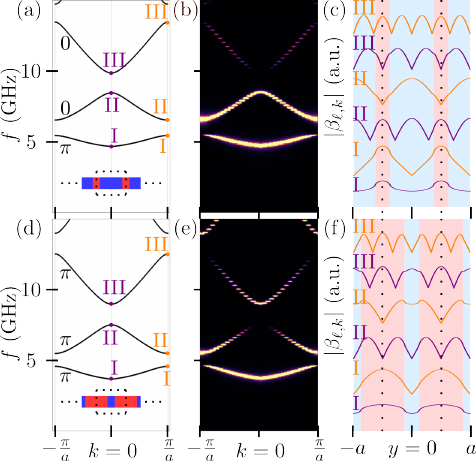}
    \caption{Dispersion relations and spin wave profiles of the bulk MCs. (a and d) Dispersion relations for ${l}/{a}=0.25$ and $0.75$ obtained by solving the LL equation. In the insets, dashed lines designate the unit cells. The purple and orange roman numbers(I-III) distinguish the magnetic field profiles for the three lowest bands at $k=0$ and $k={\pi}/{a}$ in (c) and (f). Here, $y_0$ is the center of a unit cell. (b and e) Numerically simulated dispersion relations of the corresponding MCs.}
    \label{fig2}
\end{figure}

The associated dispersion relations, response to microwave field, and magnetic field profiles are shown in Fig.~\ref{fig2}(a-c) for ${l}/{a}=0.25$ and Fig.~\ref{fig2}(d-f) for ${l}/{a}=0.75$. 
In Fig.~\ref{fig2}(a) and~\ref{fig2}(d), magnon dispersion relations at ${l}/{a}=0.75$ and $0.25$ with $y_0$ at the center of the $\text{YAIG}$ are plotted. Corresponding unit cells are illustrated by dashed lines in the insets. The Zak phase of the second band is $0$ at ${l}/{a}=0.75$ and changes to $\pi$ for ${l}/{a}=0.25$. For the second gap, the transition between topologically non-trivial and trivial MC structures occurs at ${l}/{a}\approx 0.46$ when the gap between the first and second bands closes at $k=0$. Fig.~\ref{fig2} (c) and (f) display the magnetic field profiles of the associated dispersion relations. The purple and orange lines  distinguish the magnetic field profiles at $k=0$ and $k={\pi}/{a}$, respectively. Comparison of the magnetic field profiles of the second and the third band at $k=0$ indicate that a band inversion occurred when the gap was closed.

{The calculated responses of  magnonic crystals to a spatially-uniform microwave drive are shown in Fig.~\ref{fig2}(b) and (e) and they agree with (a) and (d). The parity of the excited spin waves is revealed by the simulated band structures; even parity spin waves produce a strong signal and odd parity spin waves produce a weak signal in the dispersion relations. For instance the second band of (b) at $k=0$ is strong, and in (c), we see that second band spin wave profile at $k=0$ has even parity. On the other hand, the second band is invisible at $k=0$ in (e) and this is due to the odd parity of the excited spin waves as illustrated in (f).}

Figs. 2(b) and (e) are simulated with the micromagnetic modeling software MUMAX3\cite{MUMAX3}, with a discretized  MC of $128$ $\text{nm}$ $\times$ $4096$ $\text{nm}$ $\times$ $16$ $\text{nm}$ with the size of a single cell as $4$ $\text{nm}$ $\times$ $4$ $\text{nm}$ $\times$ $2$ $\text{nm}$ along the $\hat{x}$, $\hat{y}$ and $\hat{z}$ directions, respectively. 
We choose a damping $\alpha=3\times 10^{-5}$ that is appropriate for YIG, as the damping dependence on Al composition of YAIG is poorly known. 
A microwave field $\textbf{B}_{rf}\left(t\right)=B_{rf}{\rm sinc}\left(2\pi f_{max}\left(t-{t_{s}}/{2}\right)\right)\hat{y}$  is applied in a small localized volume at the center of the MC equivalent to 2 layers normal to $\hat{y}$ ($128$ $\text{nm}$ $\times$ $8$ $\text{nm}$ $\times$ $16$ $\text{nm}$).  $B_{rf}=2$ $\text{mT}$, $f_{max}=15\ \text{GHz}$ is the maximum frequency of the simulated dispersion relations, and $t_s=30\ \text{ns}$ is the simulation time. The simulation results are obtained as a magnetization profile at a time step $\Delta t=1/\left(2f_{max}\right)$. A two dimensional fast Fourier transform is applied to $m_+$ with respect to $y$ and $t$. For the $x$ and $z$ dependence we average over the number of  cells. 


For ${l}/{a}=0.25$ the first band has even parity at $k=0$ but odd parity at $k={\pi}/{a}$ with respect to $y_0=0$ [Fig.~\ref{fig2}(c)]. Therefore $\rho^{in}\left(ig+{\pi}/{a},{L}/{2}\right)<0$ between the first and second band. In contrast, the second band has the same parity at $k=0$ and $k={\pi}/{a}$ and  the sign does not change. {However, }$\rho^{in}\left(ig+{2\pi}/{a},{L}/{2}\right)<0$ in the gap between the second and third bands. {Similarly, }the sign of the logarithmic function is negative in the gap between the third and fourth band. 

\begin{figure}
    \centering
    \includegraphics{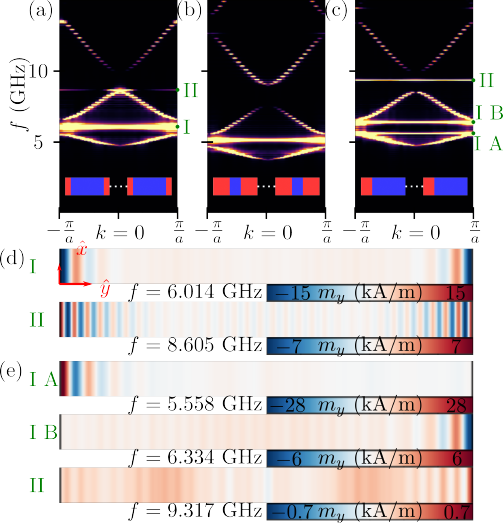}
    \caption{Simulated dispersion relations and edge states of finite MCs. (a-c) Dispersion relations of the symmetric (a and b) and asymmetric (c) finite MCs. The flat bands indicate localized excitations. The insets depict crystal terminations of the simulated systems. (d and e) Localized excitation in the MCs in (a) and (c), correspondingly. Green roman numbers are used to distinguish edge states excited at different frequencies.}
    \label{fig3}
\end{figure}

Simulations of the finite structure with inversion symmetry are done with the sinc pulse applied to a layer normal to $\hat{y}$ at each side of the edges of the MC ($128$ $\text{nm}$ $\times$ $4$ $\text{nm}$ $\times$ $16$ $\text{nm}$). This preferentially identifies signals from the edges, corresponding to flat bands in {Fig.~\ref{fig3}(a)} at $f=6.014$ $\text{GHz}$ and $8.605$ $\text{GHz}$. Furthermore, the edge state II is not found in {(b)}, which is consistent with our analysis of the Zak phases and the existence conditions for edge states.

Excitation with a simulated microwave  field $ \textbf{B}_{rf}\left(t\right)=B_{rf}\sin\left(2\pi f_{res}t\right)\hat{y}$, with  $B_{rf}=0.5$ $\text{mT}$ and
 $f_{res}$  the resonance frequency of the edge states, reveals the spin wave profiles of the edge states. Using  localized fields may produce spin wave profiles with better resolution, but might bias arbitrary bulk excitations to appear similar to edge excitations, so the excitation field is chosen to be spatially uniform. The corresponding edge state profiles are displayed in Fig.~\ref{fig3}(d). The lowest edge state appears only when the crystal is terminated with $\text{YIG}$. This defines the unit cells so that the $y_0$ is at the center of $\text{YAIG}$. The second edge state is confirmed only when ${l}/{a}>0.46$ and $\phi_{2}=0$.
In Fig.~\ref{fig3}(a), flat bands corresponding to the edge states are observed in the first and second gaps. An edge state in the third gap is not visible in Fig.~\ref{fig3}(a) due to the small amplitude of higher order excitations in the simulation.

We have also studied  finite structures without inversion symmetry corresponding to Fig.~\ref{fig1}(d). 
In Fig.~\ref{fig3}(c) and (e) a simulated dispersion relation and the edge state profiles at ${l}/{a}=0.25$ are presented. For the asymmetric MC the edge states are localized on only one side of the MC due to the broken symmetry. Compared to its symmetric counterpart, the lowest edge state is separated into two edge states that are localized on each side of the MC (Fig.~\ref{fig3}, I to I A and I B). The lower edge state is localized at the $\text{YIG}$ termination but the other edge state is at the $\text{YAIG}$ termination. Such a result can be understood as a magnonic analogue of the SSH model. In the SSH model, a perturbation at one termination causes a separation of the edge state. An edge state at the unperturbed termination is untouched, but  the frequency of the other edge state is shifted proportionally to the strength of the perturbation. On the other hand, the second edge state shows a distinct response to the broken symmetry compared to the two lowest edge states. There is only one edge state localized at the YAIG termination. This is certainly a different type of edge state compared to the lower edge state and the one observed in the SSH model.

\begin{figure}
    \centering
    \includegraphics{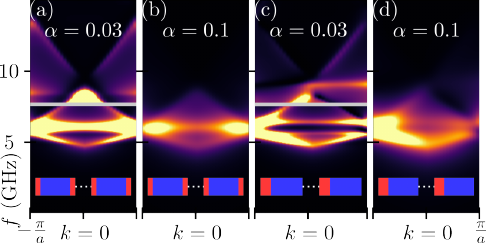}
    \caption{Simulated dispersion relations of the finite MCs with high damping constants $\alpha$. (a-b) Symmetric MCs having the same structure as Fig.~\ref{fig3} (a) and (c-d) Asymmetric MCs having the same structure as Fig.~\ref{fig3} (c). For (a) and (c), dispersion relations with different color scales are combined to make the edge state signals more visible. The gray lines distinguish the dispersion plots with different color scales.}
    \label{fig4}
\end{figure}

Finally, we comment on the robustness of the topologically protected edge states to a phenomenological damping of the spin waves. We calculate the Zak phases of the MC with $\alpha>3\times10^{-5}$ and confirm that the deviation of the Zak phase from the quantized value increases linearly with the increase of $\alpha$. Nevertheless the deviation is exceptionally  small ($\approx10^{-10}$). We also perform  micromagnetic simulations of the dispersion relations with  higher damping constants as shown in Fig.~\ref{fig4}.(a) and (b). These are symmetric MCs, identical to the MC described in Fig.~\ref{fig3}(a), but with higher damping. The edge state signals are visible even at $\alpha=0.03$. The high order edge state signal disappears for $\alpha=0.1$, but this  is due to the weak signal strength  typical of high order mode excitations. Similarly, Fig.~\ref{fig4}(c) and (d) are dispersion relations of the MC described in Fig.~\ref{fig3} (c) but with higher damping constants. All three edge state signals are visible at $\alpha=0.03$. The low order edge states are still excited even when $\alpha=0.1$ even though the signals become dimmer. These simulation results indicate that the topologically protected edge states are still observable even in highly damped MCs.

To summarize, we simulated the topological properties of  one-dimensional MCs  and confirmed  bulk-boundary correspondence,  connecting the presence of edge states to the Zak phases via the magnetic field parities at $k=0$ and $k={\pi}/{a}$. The swapping of topological character with swapped  magnetic region and the consequences of breaking inversion symmetry exhibited  similar features as the SSH model. These  Zak phases are robust;  micromagnetic simulations predict that these topologically protected edge states are observable in highly damped MCs.

\begin{acknowledgments}
    Research primarily supported as part of the Center for Energy Efficient Magnonics, an Energy Frontier Research Center funded by the US Department of Energy (DOE), Office of Science, Office of Basic Energy Sciences (BES) under Award No. DE-AC02-76SF00515 (analysis, computation, preparation for publication) and by NSF DMR-1808742 (computational framework and initial calculations). 
\end{acknowledgments}

\bibliography{apssamp}

\end{document}